\documentclass[12pt]{article}

\usepackage[english]{babel}
\usepackage{a4wide}
\usepackage{amsmath}
\usepackage{amsthm}
\usepackage{amsfonts}
\usepackage{amssymb}
\usepackage{changebar}
\usepackage{cite}
\usepackage{graphicx}

\newtheorem{prop}{Proposition}[section]

\newtheorem{theorem}{Theorem}[section]
\newtheorem{remark}[theorem]{Remark}

\DeclareMathOperator{\Tr}{Tr} \DeclareMathOperator{\im}{Im}
\DeclareMathOperator{\e}{e}

\numberwithin{equation}{section}

\begin{document}

\thispagestyle{empty} \vspace*{-80pt}
{\hspace*{\fill}} \vspace{55pt}
\begin{center}
{\Large  \textbf{Equilibrium states for the Bose gas}
  }
 \\[30pt]

{\large
    Lieselot~Vandevenne
       \footnote{KULeuven, lieselot.vandevenne@fys.kuleuven.ac.be},
    Andr\'{e}~Verbeure
       \footnote{KULeuven, andre.verbeure@fys.kuleuven.ac.be}\\
       and \\
    Valentin A.~Zagrebnov
       \footnote{U II~--~Marseille, zagrebnov@cpt.univ-mrs.fr}}
\\[30pt]
Instituut voor Theoretische Fysica, Katholieke Universiteit
Leuven, Celestijnenlaan~200D, B-3001 Leuven, Belgium\\
Universit\'e de la M\'editerran\'ee et Centre de Physique
Th\'eorique - CNRS, \\ Campus de Luminy-Case 907, F-13288
Marseille, Cedex 09, France
\\[25pt]
\end{center}
\begin{abstract}\noindent
The generating functional of the cyclic representation of the CCR
(Canonical Commutation Relations) representation for the
thermodynamic limit of the grand canonical ensemble of the free
Bose gas with attractive boundary conditions is rigorously
computed. We use it to study the condensate localization as a
function of the homothety point for the thermodynamic limit using a sequence of growing convex containers. The Kac function is
explicitly obtained proving non-equivalence of ensembles in the
condensate region in spite of the condensate density being zero locally. 
\\[15pt]

\end{abstract}

\newpage

\section{Introduction}

The interest in the phenomenon of standard Bose Einstein
Condensation (BEC) revived in recent years due to the spectacular
experimental work on Bosons in traps. We refer, e.g., to \cite
{PS} and \cite{BZ} for experimental and theoretical state of
affairs. A renewed interest in old problems connected with the
phase transition accompanying BEC is at order. The generic model
for BEC is the free Bose gas as already was pointed out by Bose
and Einstein in 1925. On the level of mathematical physics, the
understanding of the phase transition started with the well known
paper of Araki and Woods \cite{AW}, where the generating
functionals of the cyclic representations of the canonical
commutation relations corresponding to the equilibrium states of
the free Bose gas are computed for periodic boundary conditions.
Lewis and Pul\'e \cite{LP1,LP2,TVV} computed the grand canonical
equilibrium states for a set of boundary conditions including the
Dirichlet and Neumann boundary conditions but not the attractive
boundary conditions. 
They are using the Kac method. An important
consequence of their result is the explicit computation of a
non-trivial Kac density showing non-equivalence of the canonical
and grand-canonical ensembles in the condensate region. The next
result is found in \cite{BNZ}, where the same conclusion was
obtained for generalized condensations in some models of
imperfect gases with diagonal interactions.

In the present paper we complete this computation of the
equilibrium states for the free Bose gas with \textit{attractive}
boundary conditions. 
About the relevance of this type of boundary conditions, see e.g.
\cite{R,IL}. 
This model has a particular type of
condensation namely condensation in quantum states corresponding
to isolated points in the spectrum. It is well known \cite{R},
\cite{LW} that in this case the condensate is situated at the
"boundary" and \textit{not} uniformly spread out everywhere in
space. We give a precise formulation of the generating functional
in the frame of the theory of generating functionals on the CCR in
order to catch up the condensate. Finally we derive also that
there is \textit{non-equivalence} of ensembles, something which
was unclear until now because of the fact that the quantum
fluctuations show a pattern \cite{LV} completely different from
the free Bose gas with Dirichlet or Neumann boundary conditions.
The intuition behind this fact is related to a wondering
peculiarity of the free Bose gas with \textit{attractive} boundary
conditions \cite{R}, \cite{LW}.  If one takes the thermodynamic limit
using a sequence of growing convex domains with the point of homothety
at the origin of the coordinates, then  \textit{locally} the
condensation density is always equal to zero. As a byproduct, our
computations imply that the condensate is spatially situated in a
region logarithmically close to the boundary of these increasing
domains. In the present paper we take different positions of the
homothety point for cubic containers to show that the density of
this logarithmic stratum of condensate inherits also a spacial
anisotropy due to the choice of cubic containers.
Finally remark that for the rotating bucket case \cite{TVV}, one has also the effect of the condensate
being increased at the boundary. But this is an effect of large angular momentum and not of the boundary
conditions as in our case.

\section{CCR-Representations and the generating functional}

For details about the CCR algebra, we refer to \cite{BR}.

Let $\mathit{h}$ be a complex pre-Hilbert space with inner product
$(\cdot ,\cdot)$. A representation of the CCR over $\mathit{h}$ on
a Hilbert space $\mathcal{H}$ is a map $f\mapsto W(f)$ of
$\mathit{h}$ into the group $\mathcal{U}(\mathcal{H})$ of unitary
operators on a Hilbert space $\mathcal{H}$ satisfying the Weyl
relations:
\begin{equation}\label{weylrel}
W(f_1)W(f_2)=\exp\left\{-\frac{\mathrm{i}}{2}\im
(f_1,f_2)\right\}W(f_1+f_2)
\end{equation}
such that for each $f\in \mathit{h}$ the map $\lambda\mapsto W(\lambda f)$ of
$\mathbb{R}$ into $\mathcal{U(H)}$ is strongly continuous. By Stone's theorem,
this continuity condition implies the existence of self-adjoint operators
$\Phi(f)$ such that
\begin{equation}\label{weylop}
W(f)=\exp\{\mathrm{i}\Phi(f)\}\
\end{equation}
These $\Phi(f)$ are called field operators. The map
$f\mapsto\Phi(f)$ is linear over $\mathbb{R}$, but not linear
over $\mathbb{C}$. Using the $\Phi(f)$ we can now define the
creation and annihilation operators $a^\ast(f)$ and $a(f)$ for
$f\in\mathit{h}$ by:
\begin{equation}\label{creop}
a^\ast(f)=2^{-1/2}\{\Phi(f)-\mathrm{i}\Phi(\mathrm{i}f)\}
\end{equation}
\begin{equation}\label{annop}
a(f)=2^{-1/2}\{\Phi(f)+\mathrm{i}\Phi(\mathrm{i}f)\}
\end{equation}
A state on the CCR-algebra is a linear functional
$\omega:\mathit{h}\rightarrow\mathbb{C}$ with the properties:
$$\omega(A^\ast A)\geq 0,\quad\omega(\mathbf{1})=1,\qquad
\mbox{for } A \mbox{ linear combinations of the } W(f),\
f\in\mathit{h}$$ A representation $(W,\mathcal{H},\Omega)$ is
called a cyclic representation if $\Omega$ is a cyclic vector. A
vector $\Omega$ is cyclic if the set $\{W(f)\Omega\}_{f\in
\mathit{h}}$ is dense in $\mathcal{H}$. To each cyclic
representation $(W,\mathcal{H},\Omega)$ of the CCR corresponds a
generating functional $\mathbb{E}:\mathit{h}\rightarrow\mathbb{C}$
given by:
\begin{equation}\label{genfunc1}
\mathbb{E}(f)=\omega(W(f))=(\Omega,W(f)\Omega)
\end{equation}

\begin{prop}
A functional $\mathbb{E}: \mathit{h}\rightarrow \mathbb{C}$ is the
generating functional of a cyclic representation of the CCR if and
only if it satisfies the following conditions:
\begin{description}
\item[(i)]   $\mathbb{E}(0)=1$,\\
\item[(ii)]  $\forall f \in \mathit{h}: \lambda\mapsto\mathbb{E}(\lambda
f)$
is continuous ,\\
\item[(iii)] $\forall$ finite sets of complex numbers $c_1,\ldots , c_n$ and
elements $f_1,\ldots , f_n \in \mathit{h} \,:$  \\
$\sum_i\sum_j\mathbb{E}(f_i-f_j)\e^{\frac{\mathrm{i}}{2}\im(f_i,f_j)}\overline{c_i}
c_j \geq 0$.
\end{description}
\end{prop}


\section{Kac-density and equivalence of ensembles}

\subsection{Concrete setup}\label{setup}

Let $\Lambda_L^\nu = \left[-L/2,L/2\right]^\nu$ be a bounded
region in $\mathbb{R}^\nu$ with volume $V = L^\nu$. We put
$\mathit{h}_L=\mathcal{L}^2(\Lambda^\nu_L)$ for the Hilbert space
of the wave-functions in $\Lambda_L^\nu$ with the scalar product
$\left(f,g\right)_{\mathit{h}_L} := \int_{\Lambda^\nu_L}dx^\nu
\overline{f(x)}g(x)$. Then $\Lambda_L^\nu \subseteq
\Lambda_{L'}^\nu$ and $\mathit{h}_L \subseteq \mathit{h}_{L'}$
whenever $L\leq L'$ via natural imbedding.

 Let $t_L^\sigma$ be the self-adjoint extension of the
operator $-\Delta_L$ (with domain $dom(-\Delta_L)=
C_0^\infty(\Lambda_L^\nu)$) determined by the boundary conditions
$\partial_n\phi + \sigma \phi = 0 \ \mbox{on} \
\partial \Lambda_L^\nu$. Here $\partial_n$ is the
directional derivative in the direction of the outward normal $n$
to $\partial\Lambda_L^\nu$. If the parameter $\sigma \leq 0$, we
say that the boundary $\partial\Lambda_L^\nu$ is \textit{attractive}.\\
First we have to solve the one-dimensional one-body eigenvalue
problem on $\Lambda_L= \left[-L/2,L/2\right]$:
$$\left(t_L^\sigma \phi \right)(x) =\lambda \phi (x)$$ 
with boundary
conditions $(\sigma<0)$:
$$\left\{
\begin{array}{lll}
\left(\displaystyle\frac{d\phi}{dx}-\sigma\phi\right)_{x=-L/2} & = & 0 ,\\
\left(\displaystyle\frac{d\phi}{dx}+\sigma\phi\right)_{x=L/2} & =
& 0 .
\end{array} \right.$$
Due to these attractive boundary conditions, there are two
negative eigenvalues tending to the same limit $-\sigma^2$ (when
$L\rightarrow \infty$) and an infinite number of positive
eigenvalues (for $L|\sigma|>2$): $$\epsilon_L(0) < \epsilon_L(1) <
0 < \epsilon_L(2) < \epsilon_L(3) < \ldots  ,$$
$$\epsilon_L(0)=-\sigma^2- O(\e^{-L|\sigma|}) ,$$
$$\epsilon_L(1)=-\sigma^2+O(\e^{-L|\sigma|}) ,$$
\begin{equation}\label{spect}
k\geq 2:\ \ \left(\frac{(k-1)\pi}{L}\right)^2 < \epsilon_L(k) <
\left(\frac{k\pi}{L}\right)^2 .
\end{equation}
The corresponding eigenfunctions $\{\phi_k^L\}_{k \in
\mathbb{Z}_{+}}$ form a basis in $\mathit{h}_L$ and are given by
\begin{eqnarray}
\phi_0^L(x) & = &
\sqrt{\frac{2}{L}}\left(1+\frac{\sinh(L|\sigma|)}{L|\sigma|}\right)^{-1/2}
\cosh(-|\sigma|x) ,\nonumber\\ \phi_1^L(x) & = &
\sqrt{\frac{2}{L}}\left(-1+\frac{\sinh(L|\sigma|)}{L|\sigma|}\right)^{-1/2}
\sinh(-|\sigma|x) ,\nonumber\\ \phi_k^L(x)& = & \left\{
\begin{array}{ll}
\sqrt{\displaystyle\frac{2}{L}}\left(1+\displaystyle\frac{\sin
(\sqrt{\epsilon_L(k)}L)}
{\sqrt{\epsilon_L(k)}L}\right)^{-1/2}\cos(\sqrt{\epsilon_L(k)}x) ,
\qquad & \mbox{for $k$ \ even} ,\\
\sqrt{\displaystyle\frac{2}{L}}\left(1-\displaystyle\frac{\sin
(\sqrt{\epsilon_L(k)}L)}
{\sqrt{\epsilon_L(k)}L}\right)^{-1/2}\sin(\sqrt{\epsilon_L(k)}x) ,
\qquad & \mbox{for $k$ odd} .
\end{array}\right.\nonumber
\end{eqnarray}

\noindent The eigenvalues and the wave functions of the
corresponding multi-dimensional case have the form:
\begin{eqnarray}
E_L(\mathbf{k}) & = & \sum_{i=1}^\nu \epsilon_L(k_i) , \nonumber\\
\psi_{\mathbf{k}}^L(\mathbf{x}) & = & \prod_{i=1}^\nu
\phi_{k_i}^L(x_i) , \nonumber
\end{eqnarray}
where $\mathbf{k} = \{k_i\}_{i=1}^{\nu } \in \mathbb{Z}^{\nu}_{+}$
and $\mathbf{x} = \{x_i\}_{i=1}^{\nu } \in \Lambda_{L}^{\nu}$.

\subsection{Kac density}

The Kac density relates expectation values of observables in the
\textit{canonical ensemble} and those in the \textit{grand
canonical ensemble}. The canonical equilibrium state for a free
Bose gas in a cube $\Lambda_L^\nu$ of volume $V = L^\nu$ with total
particle density 
$\rho$ and inverse temperature $\beta$ is given
by
\begin{equation}\label{canensstate}
\omega_{L,\beta,\rho}^{can}(A)=\frac{\Tr_{\mathcal{H}_{L,B}^{(n)}}A^{(n)}
\e^{-\beta
T_L^{\sigma,(n)}}}{\Tr_{\mathcal{H}_{L,B}^{(n)}}\e^{-\beta
T_L^{\sigma,(n)}}},\qquad \mbox{where } n=[V\rho],\
dom(A^{(n)})\subset\mathcal{H}_L^{(n)} ,
\end{equation}
and $T_L^{\sigma,(n)}$ is the $n-$particle free Bose gas Hamiltonian
in the cube $\Lambda_L^\nu$ with boundary conditions defined by
$\sigma$. 
Now we consider the grand canonical equilibrium state at
chemical potential $\mu$ and inverse temperature $\beta$:
\begin{equation}\label{grandcanensstate}
\omega_{L,\beta,\mu}^{g.c.}(A)=\frac{\Tr_{\mathcal{F}_{L,B}}A
\exp\{-\beta(T_{L}^\sigma-\mu
N_L)\}}{\Tr_{\mathcal{F}_{L,B}}\exp\{-\beta(T_{L}^\sigma-\mu
N_L)\}},\qquad dom(A)\subset \mathcal{F}_{L,B} .
\end{equation}
Here $T_{L}^\sigma =\sum_{\mathbf{k}\in \mathbb{Z}_{+}^\nu}
E_L(\mathbf{k}) a^\ast (\psi^L_{\mathbf{k}})a(\psi^L_{\mathbf{k}})
$ is the free Bose gas Hamiltonian and $N_L=\sum_{\mathbf{k}\in
\mathbb{Z}_{+}^\nu} N_{L,\mathbf{k}}=\sum_{\mathbf{k}\in
\mathbb{Z}_{+}^\nu} a^\ast
(\psi^L_{\mathbf{k}})a(\psi^L_{\mathbf{k}})$, is the particle
number operator in $\mathcal{F}_{L,B}$, the boson Fock space over
$\mathcal{L}^2(\Lambda_L^\nu)$:
\begin{eqnarray}
\mathcal{F}_{L,B} & = &
\mathcal{F}_{B}(\mathcal{L}^2(\Lambda_L^\nu))\nonumber\\
 & = & \bigoplus_{n=0}^\infty\mathcal{H}_{L,B}^{(n)}\label{fockspace}
\end{eqnarray}
with $\mathcal{H}_{L,B}^{(n)}$ the symmetrized  $n-$particle
Hilbert space appropriate for bosons and
$\mathcal{H}_{L,B}^{(0)}=\mathbb{C}$.\\ Notice that in the
thermodynamic limit $L \rightarrow \infty$ the canonical ensemble
state $\omega_{\beta,\rho}^{can}(\cdot)$ may not coincide with the
equilibrium state of the grand canonical ensemble state
$\omega_{\beta,\overline{\mu}(\beta,\rho)}^{g.c.}(\cdot)$ for the
corresponding particle density $\rho$.  
Here $\overline{\mu}(\beta,\rho) = \lim_{L \rightarrow \infty
}\overline{\mu}_{L}(\beta,\rho)$ and $
\overline{\mu}_{L}(\beta,\rho)$ is a solution of the grand
canonical particle density equation (see also (\ref{chempotent}))
\begin{equation}\label{g.c.density}
\rho =
\omega_{L,\beta,\overline{\mu}_{L}(\beta,\rho)}^{g.c.}(N_L/V).
\end{equation}
By virtue of (\ref{fockspace}) the states (\ref{canensstate}) and
(\ref{grandcanensstate}) are related by
\begin{equation}\label{g.c./can}
\omega_{L,\beta,\mu}^{g.c.}(A) =
\int_{\mathbb{R}_+}K_{L,\beta,\mu}^{g.c.}(d\xi)\,\,\omega_{L,\beta,\xi}^{can}
(A^{\left([V \xi]\right)}),
\end{equation}
where $A^{(n)} = A \lceil  \mathcal{H}_{L,B}^{(n)}$ is a
restriction of the operator $A$ on the subspace
$\mathcal{H}_{L,B}^{(n)}$ and
\begin{equation}\label{K}
K_{L,\beta,\mu}^{g.c.}(\xi)  = \frac{\sum_{n=0}^{[V \xi
]}\exp(n\beta\mu) \Tr_{\mathcal{H}_{L,B}^{(n)}}\exp(-\beta
T_L^{(n)})}{\Tr_{\mathcal{F}_{L,B}} \exp\{-\beta(T_L-\mu N_L)\}}.
\end{equation}
For a given grand canonical density (\ref{g.c.density}), the
measure (\ref{K}) takes the form :
\begin{equation}\label{K1}
K_{L,\beta,\rho}(d\xi): =
K_{L,\beta,\overline{\mu}_{L}(\beta,\rho)}^{g.c.}(d\xi) = d\xi \,
K_{L,\beta}(\xi;\rho),
\end{equation}
and the limit $K_{\beta}(x;\rho) = \lim _{L\rightarrow\infty}
K_{L,\beta}(x;\rho)$ is known as the \textit{Kac density}, see
e.g. \cite{LP2}. If the Kac density happens to be a
$\delta$-function with support at $\rho$ , then clearly one has
(\textit{strong}) equivalence of ensembles:
\begin{equation}\label{Equiv}
\omega_{\beta,\overline{\mu}(\beta,\rho)}^{g.c.}(A) =
\omega_{\beta,\rho}^{can}(A),
\end{equation}
Otherwise there is only \textit{weak} equivalence of ensembles,
see \cite{BNZ}.

 The limit $\rho_c(\beta):= \lim_{\mu \rightarrow -\nu
\sigma^2}\lim_{L\rightarrow\infty}
\omega_{L,\beta,\mu}^{g.c.}(N_L/V)$ is the critical density for
the free Bose gas in a box with attractive boundary conditions. We
shall show that in the model the canonical and the grand canonical
ensembles are \textit{not} equivalent in the \textit{presence} of
the Bose condensate, i.e. for $\rho>\rho_c(\beta)$, or for $\beta
>\beta_c(\rho)$, where $\rho_c(\beta_c(\rho))=\rho$. 
The non-equivalence of ensembles in the case of the free Bose gas with
\textit{attractive boundaries} is not the same phenomenon as in
the case of the one with, for example Dirichlet, $\sigma = \infty
$, or Neumann, $\sigma = 0$, boundary conditions. In the case of
the attractive boundary conditions ($\sigma<0$), the condensation
phenomenon is a \textit{surface effect} (not a bulk effect as in
the free Bose gas with $\sigma = \infty $ or $\sigma = 0$): the
condensate is located near the walls, see Section 4.2.

To determine the Kac density, we have to calculate (see
(\ref{g.c./can}))
\begin{eqnarray}
\omega_{L,\beta,\overline{\mu}_{L}(\beta,\rho)}^{g.c}(W(f)) & = &
\int_{\mathbb{R}_+}K_{L,\beta,\rho}(d\xi) \,\,
\omega_{L,\beta,\xi}^{can}
(W(f))\nonumber\\
& = & \int_{\mathbb{R}_+}d\xi\ K_{L,\beta}(\xi;\rho)\
\omega_{L,\beta,\xi}^{can}(W(f)) \label{equiv}
\end{eqnarray}
for any test function $f\in C_0^{\infty}(\mathbb{R}^{\nu})$, the
$C^\infty$-functions on $\mathbb{R}^\nu$ with compact support.
Therefore we first must calculate the limit of the expectation
value of the exponential function:
\begin{equation}\label{omega1}
\omega_{\beta,\rho}^{g.c}(W(f)): =\lim_{L \rightarrow
\infty}\omega_{L,\beta,\overline{\mu}_{L}(\beta,\rho)}^{g.c}
(W(f))
\end{equation} 
with $\{\overline{\mu}_{L}(\beta,\rho)\}_{L}$ solutions of the density
equation (\ref{g.c.density}).

This is possible because the states
$\omega_{L,\beta,\mu}^{g.c}$, where $\mu < - \nu \sigma^2$, are
\textit{quasi-free states}, and these are easily obtained by using
the \textit{truncated functionals}
$\omega_{L,\beta,\rho}(\ldots)_T$, see e.g. \cite{BR}. The
functionals are defined by the recursion relations:
\begin{equation}\label{omega2}
\omega_{L,\beta,\mu}^{g.c}(A_1\ldots A_n)=\sum_{\tau \in
P_n}\prod_{J \in \tau}\omega_{L,\beta,\mu}^{g.c}(A_{j(1)},\ldots ,
A_{j(\mid J\mid)})_T
\end{equation}
for all $A_i \,(i=1,2)$ creation or annihilation operators and $n
\in \mathbb{N}$. The sum $\tau \in P_n$ is over all partitions
$\tau$ of a set of n elements into ordered subsets $J=\{j(1),
\ldots ,j(\mid J\mid)\}\in \tau$. One can verify that the
truncated functionals associated to the equilibrium states
$\omega_{L,\beta,\mu}^{g.c}$ satisfy
\begin{eqnarray}
\omega_{L,\beta,\mu}^{g.c}(a^\sharp (f))_T =
\omega_{L,\beta,\mu}^{g.c}(a^\sharp (f)) & = & 0 ,\nonumber\\
\omega_{L,\beta,\mu}^{g.c}(a^\ast(f_1),a^\ast(f_2))_T & = &
\omega_{L,\beta,\mu}^{g.c}(a(f_1),a(f_2))_T \ = \ \ 0 ,\nonumber\\
\omega_{L,\beta,\mu}^{g.c}(a^\ast(f_1),a(f_2))_T & = & \left(f_2,
\frac{1}{\e^{\beta \left(t_L^{\sigma}- \mu\right)
}-1}f_1\right)_{\mathit{h}_L},\label{2puntexpl}
\end{eqnarray}
with $f,f_1,f_2,\ldots \in \mathit{h}_L$, the space of testfunctions with support in
$\Lambda_L^\nu$, $a^\sharp = \{a\,\,
or\,\, a^\ast\}$ and $t_L^{\sigma}$ is the self-adjoint extension
of the Laplacian $-\Delta_L$ corresponding to attractive boundary
conditions $\sigma<0$ on $\partial \Lambda_L^\nu$. Then the
non-trivial two-point functions (\ref{2puntexpl}) are explicitly
given by
\begin{eqnarray}\label{2puntexpl-1}
\omega_{L,\beta,\mu}^{g.c}(a^\ast(f_1),a(f_2))_T & = &
\omega_{L,\beta,\mu}^{g.c}(a^\ast(f_1)a(f_2)) \nonumber\\ & = &
\sum_{\mathbf{k}\in
\mathbb{Z}_{+}^\nu}\overline{\hat{f}_{2}(\mathbf{k})}\hat{f}_{1}(\mathbf{k})
\frac{1}{\e^{\beta(E_L(\mathbf{k})-\mu)}-1},
\end{eqnarray}
where the transformation
$f(\mathbf{x})\mapsto\hat{f}(\mathbf{k})$ of $f \in
C_{0}^{\infty}(\mathbb{R}^{\nu})$, is now defined by
\begin{equation} \label{f-hat}
\hat{f}(\mathbf{k}):=(\psi_{\mathbf{k}}^L,f)_{\mathit{h}_L}=
\int_{\Lambda_{L}^\nu} d\mathbf{x}\
\overline{\psi_{\mathbf{k}}^L(\mathbf{x})}f(\mathbf{x}),
\end{equation}
the Fourier transforms for the basis of $t_L^\sigma$ (see section \ref{setup}).\\
Now
\begin{eqnarray}
\omega_{L,\beta,\mu}^{g.c.}(W(f)) & = &
\omega_{L,\beta,\mu}^{g.c.}(\e^{\mathrm{i}\Phi(f)})\nonumber\\
& = & \exp\sum_{n=1}^\infty
\frac{\mathrm{i}^n}{{n!}}\, \omega_{L,\beta,\rho}(\underbrace{\Phi(f),\Phi(f),\ldots,
\Phi(f)}_{\mbox{n times}})_T ,\label{omega3}
\end{eqnarray}
where the
$\omega_{L,\beta,\mu}^{g.c.}(\Phi(f),\Phi(f),\ldots,\Phi(f))_T$
are the $n-$point truncated field correlation functions. Because
of the fact that $\omega_{L,\beta,\mu}^{g.c.}$ is a quasi-free
state, only the two-point truncated correlation function is
non-vanishing, yielding:
\begin{equation}\label{omega3-1}
\omega_{L,\beta,\mu}^{g.c.}(W(f))=
\exp\left(-\frac{1}{2}\omega_{L,\beta,\mu}^{g.c.}(\Phi(f),\Phi(f))_T
\right)
\end{equation}
By virtue of (\ref{creop}) and (\ref{annop}) it can be rewritten
in terms of the creation and annihilation operators $a^\ast(f)$
and $a(f)$:
\begin{equation}\label{omega4}
\omega_{L,\beta,\mu}^{g.c.}(\Phi(f),\Phi(f))_T=\frac{1}{2}(f,f)_{\mathit{h}_L}+
\omega_{L,\beta,\mu}^{g.c.}(a^\ast(f)a(f))
\end{equation}
so that the explicit form of the generating functional
(\ref{omega3-1}) becomes:
\begin{eqnarray}
\omega_{\beta,\mu}^{g.c.}(W(f)) & = &
\lim_{L\rightarrow\infty}\omega_{L,\beta,\mu}^{g.c.}(W(f))\nonumber\\
& = & \exp\left(-\frac{1}{4}(f,f)_{\mathit{h}_L} -
\frac{1}{2}\lim_{L\rightarrow\infty}\omega_{L,\beta,\mu}^{g.c.}
(a^\ast(f)a(f))\right)\label{genfunc2}
\end{eqnarray}
A last remark about the thermodynamic limit. Notice that the
grand-canonical ensemble for the free Bose gas exists only for
$\mu < \inf spec\,(t_L^{\sigma})$. Therefore, the solution of
equation (\ref{g.c.density}) verifies the inequality
$\overline{\mu}_L(\beta,\rho)< -\nu\sigma^2$.  
Since the \textit{critical density}
\begin{equation}\label{crit.density}
\lim_{\mu \rightarrow -\nu\sigma^2}\lim_{L\rightarrow\infty}
\omega_{L,\beta,\mu}^{g.c.}\left(\frac{N_L}{L^\nu}\right) = \rho_c(\beta)
\end{equation}
for the free Bose gas with attractive boundary conditions $\sigma
< 0$  is \textit{finite} for all dimensions greater then, or equal to
one, \cite{R}, \cite{LW}, Bose-Einstein condensation occurs
for $\rho > \rho_c(\beta)$:
\begin{equation}\label{densityzeromode}
\rho_0(\beta):= \rho - \rho_c(\beta) =
\lim_{L\rightarrow\infty}2^\nu
\omega_{L,\beta,\overline{\mu}_L(\beta,\rho)}^{g.c.}\left(\frac{N_{L,\mathbf{0}}}{L^\nu}\right)
> 0,
\end{equation}
where $N_{L,\mathbf{0}}$ is the number-operator on $\mathcal{F}_{L,B}$ of
the zero mode $\mathbf{k}=\mathbf{0}$.
The factor $2^\nu$ is due to
the asymptotic degeneracy of the $\inf spec(t_L^{\sigma})= -\nu
\sigma^2 + O(\e^{-L\left|\sigma \right|})$ for $L\rightarrow\infty
$ , see Section 3.1. Notice that (\ref{densityzeromode}) implies
that the solution of (\ref{g.c.density}) for $\rho >
\rho_c(\beta)$ has the asymptotics:
\begin{equation}\label{chempotent}
\overline{\mu}_L(\beta,\rho)= -\nu \sigma^2 - \frac{2^\nu}{\beta
(\rho - \rho_c(\beta))L^{\nu}}+ o(L^{-\nu}).
\end{equation}
We use this result in the computations of the thermodynamic limit
of the generating functional below.

We conclude this section by the following statement about the
explicit form of the Kac density for the thermodynamic limit of
the free Bose gas in the \textit{cubic} box with
\textit{attractive} boundary conditions.
\begin{theorem}\label{kac}
For the free Bose gas $T_{L}^\sigma$ with attractive boundary
conditions $\sigma < 0$ the limiting Kac 
density has the form :
\begin{eqnarray}\label{kacdens}
\lefteqn{K_{\beta}(\xi;\rho)}\\ & = & \left\{
\begin{array}{ll} \delta(\xi-\rho), \!\!& \mbox{for $\rho <
\rho_{c}(\beta)$},\\ \displaystyle\frac{2^\nu \theta (\xi -
\rho_{c}(\beta))}{(2^\nu - 1)! (\rho -
\rho_{c}(\beta))}\left[\displaystyle\frac{2^\nu (\xi -
\rho_{c}(\beta))}{\rho - \rho_{c}(\beta)}\right]^{2^{\nu} -1}
\exp\left\{- \displaystyle\frac{2^\nu (\xi -
\rho_{c}(\beta))}{\rho - \rho_{c}(\beta)}\right\},\!\! & \mbox{for
$\rho \geq \rho_{c}(\beta)$}.
\end{array}\right.\nonumber
\end{eqnarray}
Here $\theta(z \leq 0)= 0$ and $\theta(z > 0)= 1$.
\end{theorem}
\noindent \textit{Proof}: By the identity (\ref{equiv}), the Kac
density $K_{\beta}(\xi;\rho)$ is 
related to the thermodynamic
limit of the characteristic function of the particle density
$N_{L}/V$ for $t\in \mathbb{R}^1$:
\begin{eqnarray}
\lim_{L\rightarrow\infty}\omega_{L,\beta,\overline{\mu}_{L}(\beta,\rho)}^{g.c}
(\exp(it N_{L}/V)) & = & \int_{\mathbb{R}_+}K_{\beta,\rho}(d\xi)
\,\, \omega_{\beta,\xi}^{can}
(\exp(it \xi))\nonumber\\
& = & \int_{\mathbb{R}_+}d\xi\ K_{\beta}(\xi;\rho)\exp(it \xi),
\label{equiv-1}
\end{eqnarray}
To calculate the limit in the left-hand
side of (\ref{equiv-1}), we use that the state
$\omega_{L,\beta,\overline{\mu}_{L}(\beta,\rho)}^{g.c}(\cdot)$ is
quasi-free. Then
\begin{eqnarray}
\lefteqn{\omega_{L,\beta,\overline{\mu}_{L}(\beta,\rho)}^{g.c}
(\exp(it N_{L}/V))}\nonumber\\ & = & \prod_{\mathbf{k}\in
\mathbb{Z}_{+}^\nu}\left\{\frac{1- \exp\left[-\beta
(E_{L}(\mathbf{k})- \overline{\mu}_{L}(\beta,\rho))\right]}{1-
\exp\left[-\beta (E_{L}(\mathbf{k})-
\overline{\mu}_{L}(\beta,\rho)- it/\beta L^\nu )\right]}\right\}.
\label{charact}
\end{eqnarray}
Since the $2^\nu$ lowest energy-levels, i.e. the levels for which $\mathbf{k}\in
\mathbb{K}_{\leq 2^\nu}=\{\mathbf{k}\in\mathbb{Z}_+^\nu:k_i=0,1;\, i=1,\ldots ,\nu\}$ are
exponentially degenerated when $L\rightarrow\infty$ :
$E_{L}(\mathbf{k}\in \mathbb{K}_{\leq 2^\nu}) = -\nu \sigma^2 +
O(\e^{-L\left|\sigma \right|})$, by virtue of (\ref{chempotent})
and (\ref{charact}) we get that
\begin{eqnarray}
\lefteqn{\lim_{L\rightarrow\infty}\omega_{L,\beta,\overline{\mu}_{L}(\beta,\rho)}^{g.c}
(\exp(it N_{L}/V))}\nonumber\\ & = & \left\{
\begin{array}{ll} \exp(it \rho), \!\!& \mbox{for $\rho \leq
\rho_{c}(\beta)$},\\ \left[1 - it 2^{-\nu}(\rho -
\rho_{c}(\beta))\right]^{-2^{\nu}} \exp(it\rho_{c}(\beta)) ,\!\! &
\mbox{for $\rho
> \rho_{c}(\beta)$}.
\end{array}\right.\label{charact-1}
\end{eqnarray}
Therefore, by (\ref{equiv-1}), the Kac density
(\ref{kacdens}) is the Fourier transformation of the right-hand
side of (\ref{charact-1}). \hfill $\Box$\\

\section{The generating functional}
\subsection{Condensate and generating functional}

We are interested in the thermodynamic limit of the generating
functional $\omega_{\beta,\rho}^{g.c}(W(f))=
\lim_{L\rightarrow\infty}\omega_{L,\beta,\overline{\mu}_L(\beta,\rho)}^{g.c}(W(f))$
for any $f\in C_0^{\infty}(\mathbb{R}^{\nu})$. To this end we
choose the box $\Lambda_L^\nu$ with $L$ large enough such that the
$\Lambda_f : = supp\, (f)$ is contained in $\Lambda_L^\nu$.
We consider here the generating functional
$\omega_{\beta,\rho}^{g.c}(W(f))$ for $f$ an element in
$C_0^{\infty}(\mathbb{R}^{\nu})$.
\begin{theorem}\label{stell1}
The generating functional $\omega_{\beta,\rho}^{g.c}(W(f))$ on
$C_0^{\infty}(\mathbb{R}^{\nu})$ is given by:
\begin{eqnarray} \label{genfunc3}
\omega_{\beta,\rho}^{g.c}(W(f)) & = &
\exp\left(-\frac{1}{4}(f,f)\right)\exp\left(-\frac{1}{2}(f,
g_{\sigma}(\beta,\rho) f)\right),
\end{eqnarray}
with operator $g_{\sigma}(\beta,\rho)$ on
$\mathcal{L}^2(\mathbb{R}^\nu)$ defined by
\begin{eqnarray}\label{green}
(g_{\sigma}(\beta,\rho)f)(\mathbf{x}) & = &
\int_{\mathbb{R}^\nu}d\mathbf{y}\
G_{\sigma}(\beta,\rho)(\|\mathbf{x}-\mathbf{y}\|)f(\mathbf{y}),\nonumber\\
G_{\sigma}(\beta,\rho)(r) & = &
(4\pi\beta)^{-\nu/2}\sum_{n=1}^\infty \e^{-r^2/4n\beta}\
\frac{\e^{-n\beta\overline{\mu}(\beta,\rho)}}{n^{\nu/2}},
\end{eqnarray}
where $\overline{\mu}(\beta,\rho) < - \nu
\sigma^2$ for $\rho < \rho_c(\beta)$ and $\overline{\mu}(\beta,\rho) = -
\nu \sigma^2$ for $\rho \geq \rho_c(\beta)$are limiting solutions of the grand canonical
density equation (\ref{g.c.density}).
\end{theorem}
\noindent \textit{Proof}: In order to determine the generating
functional $\omega_{\beta,\rho}^{g.c}(W(f))$, we have to compute
$\lim_{L\rightarrow\infty}\omega_{L,\beta,\overline{\mu}_L(\beta,\rho)}^{g.c}
(a^\ast(f)a(f))$, see (\ref{genfunc2}). Since the attractive
boundary conditions $\sigma < 0$ create a gap in the spectrum
$spect\, (t_{L}^{\sigma})$, and  
respectively in $spect
\,(T_{L}^{\sigma})$, the calculations need a separation of the
negative eigenvalues from the positive part of the spectrum.

We consider first the one-dimensional case, when there are only
two negative eigenvalues tending to $-\sigma^2$ for
$L\rightarrow\infty$, see Section~\ref{setup}. By virtue of
(\ref{2puntexpl-1}) one gets for a given $f\in
C_0^{\infty}(\mathbb{R}^{1})$ that
\begin{eqnarray}
\lefteqn{\lim_{L\rightarrow\infty}
\omega_{L,\beta,\overline{\mu}_L(\beta,\rho)}^{g.c}(a^\ast(f)a(f))}\nonumber\\
& = & \lim_{L\rightarrow\infty}\sum_{k \in \mathbb{Z}_{+}^{1}}
|\hat f(k)|^2 \frac{1}{\e^{\beta(\epsilon_{L}(k)-
\overline{\mu}_L(\beta,\rho))}-1} \nonumber\\ & = &
\lim_{L\rightarrow\infty}\left(|\hat f(0)|^2
\frac{1}{\e^{\beta(\epsilon_{L}(0)-
\overline{\mu}_L(\beta,\rho))}-1}  + |\hat f(1)|^2
\frac{1}{\e^{\beta(\epsilon_{L}(1)-
\overline{\mu}_L(\beta,\rho))}-1}\right.\nonumber\\ & & + \left.
\sum_{n=1}^\infty
\e^{n\beta\overline{\mu}_L(\beta,\rho)}\sum_{k=2}^\infty
\e^{-n\beta\epsilon_L(k)} |\hat
f(k)|^2\right)\label{2puntfunc1dim}
\end{eqnarray}
\noindent As mentioned before, we choose $\Lambda_L$ large enough
such that $supp\,(f)=\Lambda_f$ is contained in $\Lambda_L$. Then
one estimates that $|\hat{f}(0)|^2$ has an asymptotics of the
order of $O(\e^{-L|\sigma|})$ for large $L$ since
\begin{eqnarray}
|\hat{f}(0)|^2 & = & \left| \int_{\Lambda_f}dx f(x)
\overline{\phi_0^L(x)}\right|^2\nonumber\\
& = & \frac{2}{L}\left(1+\frac{\sinh(L|\sigma|)}
{L|\sigma|}\right)^{-1}\left|\int_{\Lambda_f}dx f(x)
\cosh(-|\sigma|x)\right|^2\nonumber\\  & = &
4|\sigma|\e^{-L|\sigma|}\left|\int_{\Lambda_f}dx
f(x)\cosh(-|\sigma|x)\right|^2 + o (\e^{-L|\sigma|})
\label{eerste2}
\end{eqnarray} The
integral in the last expression is independent of $L$, because
$supp\, (f)$ is finite and inside the box $\Lambda_L$. Similarly
one gets for $L\rightarrow\infty$ that
\begin{equation}\label{eerste2-1}
|\hat{f}(1)|^2 = 4|\sigma|\e^{-L|\sigma|}\left|\int_{\Lambda_f}dx
f(x)\sinh(-|\sigma|x)\right|^2 + o (\e^{-L|\sigma|})
\end{equation}

\noindent Consider now the co\"efficients of $|\hat{f}_L(0)|^2$
and of $|\hat{f}_L(1)|^2$ in (\ref{2puntfunc1dim}). If $\rho <
\rho_{c}(\beta)$, then $\overline{\mu}(\beta,\rho) < - \sigma^2$,
i.e. $\epsilon_L(0)-\overline{\mu}_{L}(\beta,\rho) > 0$ for large
$L$. Therefore, by virtue of (\ref{2puntfunc1dim}),
(\ref{eerste2}) and (\ref{eerste2-1}), both of those terms are of
the order $O (\e^{-L|\sigma|})$ for large $L$. If $\rho \geq
\rho_{c}(\beta)$, then $\overline{\mu}_{L}(\beta,\rho) = -
\sigma^2 + O(L^{-1})$, and one gets for large $L$:
\begin{eqnarray}
\left\{|\hat{f}(0)|^2
\frac{1}{\e^{\beta(\epsilon_L(0)-\overline{\mu}_{L}(\beta,\rho))}-1}
+ |\hat{f}(1)|^2
\frac{1}{\e^{\beta(\epsilon_L(1)-\overline{\mu}_{L}(\beta,\rho))}
-1}\right\} \nonumber\\ \simeq
\frac{1}{2}\rho_0(\beta)L\left\{|\hat{f}(0)|^2 +
|\hat{f}(1)|^2\right\}, \label{eerste1}
\end{eqnarray}
where $\rho_0(\beta)=\rho-\rho_c(\beta)$ is the condensate
density. Therefore, again by virtue of (\ref{eerste2}) and
(\ref{eerste2-1}), these terms vanish in the limit
$L\rightarrow\infty$.

 Consider now the last term in the limit
(\ref{2puntfunc1dim}). By virtue of (\ref{f-hat}) for $\nu=1$ (see
Section 3.1) we can represent the sum over $k \geq 2$ in the
following explicit form:
\begin{eqnarray}
&& \frac{2}{L}\sum_{\left\{k\geq2 :
\,\,even\right\}}\left(1+\displaystyle\frac{\sin
(\sqrt{\epsilon_L(k)}L)}
{\sqrt{\epsilon_L(k)}L}\right)^{-1}\overline{\hat{F}_{L\,even}(k)}\hat{F}_{L\,even}(k)
\e^{-s\epsilon_L(k)} + \nonumber\\
&& \frac{2}{L}\sum_{\left\{k\geq2 :
\,\,odd\right\}}\left(1-\displaystyle\frac{\sin
(\sqrt{\epsilon_L(k)}L)}
{\sqrt{\epsilon_L(k)}L}\right)^{-1}\overline{\hat{F}_{L\,odd}(k)}
\hat{F}_{L\,odd}(k)\e^{-s\epsilon_L(k)},\label{last-1}
\end{eqnarray}
where $s=n\beta$  and
\begin{equation}\label{last-11}
\hat{F}_{L\,even}(k) : = \int_{\Lambda_f}dx
\cos(\sqrt{\epsilon_L(k)}x) f(x),\,\, \hat{F}_{L\,odd}(k) : =
\int_{\Lambda_f}dx \sin(\sqrt{\epsilon_L(k)}x) f(x).
\end{equation}
Since the spectrum $\left\{\epsilon_L(k)\right\}_{k\geq2}$
verifies the conditions (\ref{spect}) and $f\in
C_0^{\infty}(\mathbb{R}^{1})$, the first and the second series of terms in
(\ref{last-1}) are Darboux-Riemann sums for the corresponding
integrals:
\begin{eqnarray}
\lefteqn{\lim_{L\rightarrow\infty}\left\{\frac{2}{L}\sum_{\left\{k\geq2
: \,\,even\right\}}\left(1+\displaystyle\frac{\sin
(\sqrt{\epsilon_L(k)}L)}
{\sqrt{\epsilon_L(k)}L}\right)^{-1}\overline{\hat{F}_{L\,even}(k)}\,
\hat{F}_{L\,even}(k) \e^{-s\epsilon_L(k)} \right.}\nonumber\\ & +
& \left.\frac{2}{L}\sum_{\left\{k\geq2 :
\,\,odd\right\}}\left(1-\displaystyle\frac{\sin
(\sqrt{\epsilon_L(k)}L)}
{\sqrt{\epsilon_L(k)}L}\right)^{-1}\overline{\hat{F}_{L\,odd}(k)}\,
\hat{F}_{L\,odd}(k)\e^{-s\epsilon_L(k)}\right\}\nonumber\\ & = &
\frac{1}{\pi}\int_{0}^{\infty}dk\,\,
\overline{\mbox{Re}\left(\e^{ik\,\cdot},f\right)}_{\mathit{h}_L}\,\,\,
\mbox{Re}\left(\e^{ik\,\cdot},f\right)_{\mathit{h}_L}\e^{-sk^2}\nonumber\\
& & \ + \ \frac{1}{\pi}\int_{0}^{\infty}dk\,\,
\overline{\mbox{Im}\left(\e^{ik\,\cdot},f\right)}_{\mathit{h}_L}\,\,\,
\mbox{Im}\left(\e^{ik\,\cdot},f\right)_{\mathit{h}_L}\e^{-sk^2}.\label{last-2}
\end{eqnarray}
The last expression of (\ref{last-2}) yields:
\begin{eqnarray}\label{last-3}
\lefteqn{\frac{1}{2\pi}\int_{-\infty}^{\infty}dk\,\,
\overline{\left(\e^{ik\,\cdot},f\right)}_{\mathit{h}_L}\,
\left(\e^{ik\,\cdot},f\right)_{\mathit{h}_L}\e^{-sk^2}}\nonumber\\
& = & (4\pi s)^{-1/2}\int_{\mathbb{R}^1} dx \int_{\mathbb{R}^1} dy
\overline{f(x)}f(y) \exp\left\{-\frac{|x-y|^2}{4 s}\right\}.
\end{eqnarray}
Finally, taking into account (\ref{eerste2})--(\ref{eerste1}),
(\ref{last-3}), and the fact that $\overline{\mu}(\beta,\rho)\leq
-\sigma^2 < 0 $, we get for the limit (\ref{2puntfunc1dim}) in the
one-dimensional case:
\begin{equation} \label{corr-1}
\omega_{\beta,\rho}^{g.c}(a^\ast(f)a(f))= \left(f, g_{\sigma, \nu
= 1}(\beta, \rho) f\right)_{\mathit{h}_L},
\end{equation}
where $g_{\sigma, \nu = 1}(\beta, \rho)$ is the integral operator
on $\mathcal{L}^2 (\mathbb{R}^1)$ defined by
\begin{eqnarray}
(g_{\sigma, \nu = 1}(\beta,\rho) f)(x) & = & \int_{\mathbb{R}^1}dy\ G_{\sigma, \nu
= 1}(\beta,\rho)(\left|x-y\right|)f(y)\,\,,\nonumber\\
G_{\sigma, \nu = 1}(\beta,\rho)(r) & = &
(4\pi\beta)^{-1/2}\sum_{n=1}^\infty \e^{-r^2/4n\beta}\
\frac{\e^{-n\beta\overline{\mu}(\beta,\rho)}}{n^{1/2}}\,\,.\nonumber
\end{eqnarray}

\noindent Using the results for the one-dimensional case, one
computes the two-point correlation function
$\omega_{\beta,\rho}^{g.c}(a^\ast(f)a(f))$  in the
$\nu$-dimensional case.  Since the first $2^\nu$ wave functions
$\left\{\psi_{\mathbf{k}}^{L}(
\mathbf{x})\right\}_{\mathbf{k}\in\mathbb{Z}_{+}^{\nu}}$ have the
same exponential behaviour as in the one-dimensional case and
since $\inf_{L}\, spect (t_{L}^\sigma) = - \nu \sigma^2$, see
Section 3.1, we get:
\begin{eqnarray}
\lefteqn{\lim_{L\rightarrow\infty}\omega_{L,\beta,\overline{\mu}(\beta,\rho)}^{g.c}
(a^\ast(f)a(f))}\nonumber\\ & = &
\lim_{L\rightarrow\infty}\sum_{\mathbf{k} \in \mathbb{Z}_{+}^\nu}|
\hat f(\mathbf{k})|^2 \frac{1}{\e^{\beta(E_L(\mathbf{k})-
\overline{\mu}_{L}(\beta,\rho))}-1}\nonumber\\ & = &
\sum_{n=1}^\infty \e^{-n\beta\nu\overline{\mu}(\beta,\rho)}(4\pi
n\beta)^{-\nu/2}\int_{\mathbb{R}^\nu} d\mathbf{x}
\int_{\mathbb{R}^\nu}d\mathbf y \overline{f(\mathbf{x})}f(\mathbf
y)\exp\left\{-\frac{\|\mathbf{x} -
\mathbf{y}\|^2}{4n\beta}\right\} ,\nonumber
\end{eqnarray}
which implies (\ref{green}). By virtue of (\ref{genfunc2}) this
finishes the proof
of (\ref{genfunc3}) and of the theorem for any particle density $\rho$.
\hfill $\Box$\\

\noindent Theorem \ref{stell1} tells us that the condensate is not
traceable by considering only strictly local observables. 
The characteristic functional on the CCR-C$^\ast$-algebra of
quasi-local observables coincides with the one without condensate.
The reason for this is that the condensate is not homogeneous but
located in the vicinity of the container boundary.

In order to catch up the presence of the condensate or to get a
complete picture of the system, one has to extend the algebra of
observables to the weak closure of the CCR-C$^\ast$-algebra with
respect to the limit Gibbs states. In the next paragraph we
compute the limit functional on the relevant non-localised
observables, and obtain a complete picture yielding the existence
of sufficiently many fields in the representation of any
w$^\ast$-limit point of Gibbs states as $L$ tends to infinity. In
fact our strategy will be to make a relevant choice of the
\textit{homothety point} for the thermodynamic limit of convex
containers, in order to catch up the condensate.

Above and below we considered only the easy shape container limit,
namely cubic boxes. Because of the particular inhomogeneous
spreading of the condensate in the neighbourhood of the box
boundary, it is clear that this thermodynamic limit treatment can
be very much shape dependent. In this paper we do not enter into
the details of this specific problem.

\subsection{Condensate localization}

\begin{remark} It sounds curious that in spite of the non-zero condensate
density for $\rho > \rho_c(\beta)$ , (\ref{densityzeromode}),
there is no trace of it in the generating functional
(\ref{genfunc3}). This is in contrast to the Kac density
(\ref{kacdens}), which explicitly depends on the condensate
density $\rho - \rho_c(\beta)$. To understand this difference one
has to take into account that (\ref{genfunc3}) is localized on
the support of the function $f\in C_0^{\infty}(\mathbb{R}^{\nu})$
whereas the Kac density is a global function, depending on the condensate even
if it is localized at
"infinity", sticked to the attractive boundaries.
\end{remark}
\noindent In order to make this statement rigorous we start first with the
one-dimensional case. Let the function $f\in
C_0^{\infty}(\mathbb{R}^{1})$ be such that
$supp(f)=(-\delta,\delta)\subset\left(-L/2,L/2\right)$ and $\delta
< (\ln L)/2|\sigma|$. Consider its shift over a distance
$\gamma_{L}(\sigma):= {L}/{2}-(2|\sigma|)^{-1}\ln L $:
\begin{equation}\label{shift}
f_{\tau_{\gamma_L(\sigma)}}(x) \equiv
(\tau_{\gamma_L(\sigma)}f)(x):= f\left(x -
[{L}/{2}- (2|\sigma|)^{-1}\ln L] \right)
\end{equation}
Then $f_{\tau_{\gamma_L(\sigma)}}\in
C_{0}^{\infty}(-L/2,L/2)$.

 To get the generating functional we compute now the
limit of the corresponding two-point function (\ref{genfunc2}):
\begin{eqnarray}
\lefteqn{\lim_{L\rightarrow\infty}\omega_{L,\beta,\overline{\mu}_L(\beta,\rho)}^{g.\,c.}
(a^\ast(\tau_{\gamma_L(\sigma)} f)a(\tau_{\gamma_L(\sigma)}
f))}\nonumber\\ & = & \lim_{L\rightarrow\infty} \sum_{k=0}^\infty
|\hat f_{\tau_{\gamma_L(\sigma)}}(k)|^2
\frac{1}{\e^{\beta(\epsilon_L(k)-
\overline{\mu}_{L}(\beta,\rho))}-1}\nonumber\\ & = &
\lim_{L\rightarrow\infty}\left(|\hat
f_{\tau_{\gamma_L(\sigma)}}(0)|^2
\frac{1}{\e^{\beta(\epsilon_L(0)-
\overline{\mu}_{L}(\beta,\rho))}-1}\right.\nonumber\\ & & \ +
\left.|\hat f_{\tau_{\gamma_L(\sigma)}}(1)|^2
\frac{1}{\e^{\beta(\epsilon_L(1)-
\overline{\mu}_{L}(\beta,\rho))}-1} + \sum_{k=2}^\infty |\hat
f_{\tau_{\left[\gamma_L(\sigma)\right]}}(k)|^2
\frac{1}{\e^{\beta(\epsilon_L(k)-
\overline{\mu}_{L}(\beta,\rho))}-1}\right).\label{tweepunt2}
\end{eqnarray}
\begin{remark} Notice that in contrast to
(\ref{2puntfunc1dim}), the shift (\ref{shift}) corresponds simply
to the choice of a \textit{new point of homothety} for the
thermodynamic limit (\ref{tweepunt2}). In (\ref{2puntfunc1dim}),
the point of homothety coincides with the origin of coordinates
$x=0$, whereas in (\ref{tweepunt2}) this point is ${L}/{2}-
(2|\sigma|)^{-1}\ln L$.
\end{remark}
\noindent Now, and in contrast to (\ref{eerste2}),
$|\hat{f}_{\tau_{\gamma_L(\sigma)}}(0)|^2$ goes like
$L^{-1}$ for large $L$. Indeed,
\begin{eqnarray}
|\hat{f}_{\tau_{\gamma_L(\sigma)}}(0)|^2 & = &
\left|\int_{\gamma_L(\sigma) - \delta}^{\gamma_L(\sigma)+\delta}
dx (\tau_{\gamma_L(\sigma)}f)(x)
\phi_0^L(x)\right|^2\nonumber\\
& = & |\sigma|L^{-1}\left|\int_{-\delta}^\delta dx
f(x)\e^{|\sigma|x}\right|^2 + o(L^{-1})\nonumber
\end{eqnarray}
Remark that for $\rho > \rho_c(\beta)$ the first term in
(\ref{tweepunt2}) remains now \textit{finite} in the limit
$L\rightarrow\infty$. Taking into account (\ref{densityzeromode})
and (\ref{chempotent}) one gets:
\begin{eqnarray}
\lim_{L\rightarrow\infty}|\hat{f}_{\tau_{\gamma_L(\sigma)}}(0)|^2
\frac{1}{\e^{\beta(\epsilon_L(0)-
\overline{\mu}_{L}(\beta,\rho))}-1}& = &
\frac{\rho_0(\beta,\rho)}{2}|\sigma|\left| \int_{-\delta}^\delta
dx
 f(x)\e^{|\sigma|x}\right|^2. \label{term1}
\end{eqnarray}
The same reasoning for the second term in formula
(\ref{tweepunt2}) gives a similar result:
\begin{eqnarray}
\lim_{L\rightarrow\infty}|\hat{f}_{\tau_{\gamma_L(\sigma)}}
(1)|^2\frac{1}{\e^{\beta(\epsilon_L(1)-
\overline{\mu}_{L}(\beta,\rho))}-1} & = &
\frac{\rho_0(\beta,\rho)}{2} |\sigma| \left|\int_{-\delta}^\delta
dx f(x) \e^{|\sigma|x}\right|^2\label{term2}.
\end{eqnarray}
By the same computations as used in the proof of
Theorem~\ref{stell1}, the third term in (\ref{tweepunt2}) yields
for $\rho > \rho_c(\beta)$:
\begin{eqnarray}
\lefteqn{ \lim_{L\rightarrow\infty}\sum_{k=2}^\infty |\hat
f_{\tau_{\gamma_L(\sigma)}}(k)|^2
\frac{1}{\e^{\beta(\epsilon_L(k)-
\overline{\mu}_{L}(\beta,\rho))}-1}\ }\nonumber\\ & = &
\sum_{n=1}^\infty \e^{-n\beta\sigma^2}(4\pi
n\beta)^{-1/2}\int_{\mathbb{R}^1} dx \int_{\mathbb{R}^1} dy
f(x)\overline{f(y)}
\exp\left\{-\frac{|x-y|^2}{4n\beta}\right\}.\label{term3}
\end{eqnarray}
Hence the two-point function for the one-dimensional problem
becomes:
\begin{eqnarray}\label{one-dim}
\lefteqn{\lim_{L\rightarrow\infty}\omega_{L,\beta,\overline{\mu}_L(\beta,\rho)}^{g.\,c.}
(a^\ast(\tau_{\gamma_L(\sigma)}f)a(\tau_{\gamma_L(\sigma)}f))}\nonumber\\
& = & \rho_0(\beta,\rho)|\sigma|\left|\int_{\mathbb{R}^1} dx
f(x)\e^{|\sigma|x}\right|^2 + \left(f, g_{\sigma, \nu = 1}(\beta,
\rho) f\right),
\end{eqnarray}
see (\ref{corr-1}) for the definition of the operator $g_{\sigma,
\nu = 1}(\beta, \rho)$.

It is evident that one gets the same result for the shift of
$supp(f)=(-\delta,\delta)$ over a distance $-\gamma_L(\sigma) =
-{L}/{2}+(2|\sigma|)^{-1}\ln L$, i.e.:
\begin{eqnarray}\label{one-dim2}
\lefteqn{\lim_{L\rightarrow\infty}\omega_{L,\beta,\overline{\mu}_L(\beta,\rho)}^{g.\,c.}
(a^\ast(\tau_{\pm\gamma_L(\sigma)}
f)a(\tau_{\pm\gamma_L(\sigma)}f))}\nonumber\\ & = &
\rho_0(\beta,\rho)|\sigma|\left|\int_{\mathbb{R}^1}dx
f(x)\e^{\pm|\sigma|x}\right|^2 + \left(f, g_{\sigma, \nu =
1}(\beta, \rho) f\right),
\end{eqnarray}
where
\begin{equation}\label{shift-1pm}
(\tau_{\pm\gamma_L(\sigma)}f)(x): =  f\left(x \mp
\left[{L}/{2}- (2|\sigma|)^{-1}\ln L\right] \right).
\end{equation} 
Therefore, taking the thermodynamic limit $L\rightarrow\infty$ at
one of the homothety points $\pm\gamma_L(\sigma)$, we get that the
generating functional depends on the Bose-condensate density for
$\rho \geq \rho_{c}(\beta)$:
\begin{eqnarray}\label{one-dim3}
\lefteqn{\lim_{L\rightarrow\infty}\omega_{L,\beta,\overline{\mu}_L(\beta,\rho)}^{g.\,c.}
\left(W(\tau_{\pm\gamma_L(\sigma)}f)\right)}\nonumber\\ & = &
\exp\left(-\frac{1}{4}(f,f)\right)\exp\left(-\frac{1}{2}C^{\pm}
_{\sigma, \nu =1}(f) - \frac{1}{2}(f,g_{\sigma, \nu =1} f)\right),
\end{eqnarray}
where, by virtue of (\ref{one-dim2}), one has
\begin{equation}\label{one-dim4}
C^{\pm}_{\sigma,\nu =1}(f) = \rho_0(\beta,\rho) |\sigma|
\left|\int_{\mathbb{R}^1}d{x} f({x})\e^{\pm |\sigma|x}\right|^2.
\end{equation}
\begin{remark}\label{log-tuning1}
Notice that this result is due to a fine (logarithmic) tuning of
the position of the homothety points $\pm\gamma_L(\sigma)$.
Indeed, take $\pm\gamma_L(a \sigma )$, for $0< a < 1$, i.e.  
the homothety points are \textit{more distant} from the boundary $\pm L/2$. 
Taking into account the explicit form of the eigenfunctions
for $k = 0,1$ one finds that now $|\hat{f}_{\tau_{\pm\gamma_L(a \sigma)}}
(k = 0,1)|^2$ goes for large $L$ like
$L^{-1/a}$. This implies that both limits (\ref{term1}) and
(\ref{term2}), and hence (\ref{one-dim4}), vanish. So, the
generating functional (\ref{one-dim3}) has the same form as for
thermodynamic limit with the homothety point at the
\textit{origin}. In contrast to that, the choice $1 < a $ means
that the homothety points are \textit{closer} to the boundaries
$\pm L/2$. Then $|\hat{f}_{\tau_{\pm \gamma_L(a\sigma)}}(k = 0,1)|^2$ goes
slower then $L^{-1}$. This
implies that both limits (\ref{term1}) and (\ref{term2}), and
hence (\ref{one-dim4}), becomes infinite. So, for $\rho \geq
\rho_{c}(\beta)$ the generating functional (\ref{one-dim3}) is
zero, whereas for $\rho < \rho_{c}(\beta)$ it is nontrivial
with $C^{\pm}_{\sigma,\nu =1}(f) = 0$.
\end{remark}

\noindent To interpret these results, consider the \textit{local}
particle density:
\begin{equation}\label{loc-dens1}
\omega_{L,\beta,\overline{\mu}_L(\beta,\rho)}^{g.\,c.}
\left(a^\ast(x)a(x) \right) = \sum_{k \in \mathbb{Z}_{+}^{\nu=1}}
\frac{|\phi_k^L(x)|^2}{\e^{\beta(\epsilon_L(k)-
\overline{\mu}_{L}(\beta,\rho))}-1}.
\end{equation}
Here $a(x)$ is the Bose-field operator such that $a(f) =
\int_{\mathbb{R}^{\nu=1}}dx \overline{f(x)}a(x)$ for $f \in
C_0^\infty(\Lambda_L^{\nu =1})$ and $N(x) = a^\ast(x)a(x)$ is the
local number operator, cf. (\ref{2puntexpl-1}). Then by
(\ref{g.c.density}) and (\ref{2puntexpl-1}), the global density
is:
\begin{equation}\label{glob-dens}
\omega_{L,\beta,\overline{\mu}_{L}(\beta,\rho)}^{g.\,c.}\left(\frac{N_L}{L}\right)=
\frac{1}{L}\int_{\Lambda_{L}^{\nu = 1}}\, dx\,
\omega_{L,\beta,\overline{\mu}_L(\beta,\rho)}^{g.\,c.}
\left(a^\ast(x)a(x) \right).
\end{equation}

\noindent Consider the thermodynamic limit of the \textit{local}
particle density at the \textit{origin} of the coordinates $x=0$.
Taking into account the explicit form of the eigenfunctions, one
gets that
\begin{equation}\label{loc-dens2-1}
\rho(\beta,\rho;\, x=0) : =
\lim_{L\rightarrow\infty}\omega_{L,\beta,\overline{\mu}_L(\beta,\rho)}^{g.\,c.}
\left(a^\ast(x=0)a(x=0) \right) =
\frac{1}{\pi}\int_{\mathbb{R}_{+}^1}\,dk \frac{1}{\e^{\beta (k^2-
\overline{\mu}(\beta,\rho))}-1}
\end{equation}
for $\rho < \rho_{c}(\beta)$, and
\begin{equation}\label{loc-dens2}
\rho(\beta,\rho;\, x=0)=
\lim_{L\rightarrow\infty}\omega_{L,\beta,\overline{\mu}_L(\beta,\rho)}^{g.\,c.}
\left(a^\ast(x=0)a(x=0) \right) =
\frac{1}{\pi}\int_{\mathbb{R}_{+}^1}\,dk \frac{1}{\e^{\beta (k^2 +
\sigma^2)}-1} = \rho_{c}(\beta)
\end{equation}
for $\rho \geq \rho_{c}(\beta)$ by (\ref{chempotent}). 
By inspection of (\ref{loc-dens2-1}) and (\ref{loc-dens2}) based on the explicit
formulae for the eigenfunctions one readily gets that
\begin{equation}\label{loc-dens2+1}
\rho(\beta,\rho;\,x)= \rho(\beta,\rho;\,x=0)
\end{equation}
for any $x$ in a \textit{bounded} domain $D$, containing the origin of the
coordinates $x=0$.
In particular we get that the limiting \textit{local} density for
$x \in D$ corresponding to the first two modes ($k = 0,1$) is
\begin{equation}\label{loc-dens-cond1}
\rho_0(\beta,\rho;\,x) := \lim_{L\rightarrow\infty} \sum_{k =
0,1} \frac{|\phi_k^L(x=0)|^2}{\e^{\beta(\epsilon_L(k)-
\overline{\mu}_{L}(\beta,\rho))}-1} = 0 .
\end{equation}
On the other hand, the \textit{global} Bose-Einstein condensation
density (\ref{densityzeromode}) is also related exactly to these two
modes:
\begin{equation}\label{densityzeromode1}
\rho_0(\beta,\rho)= \lim_{L\rightarrow\infty}\frac{1}{L}\sum_{k =
0,1 }\frac{1}{\e^{\beta(\epsilon_L(k)-
\overline{\mu}_{L}(\beta,\rho))}-1}= \rho - \rho_{c}(\beta) > 0,
\end{equation}
which is not present in (\ref{loc-dens2}).

Consider now the local density of the Bose-Einstein condensation
(\ref{loc-dens-cond1}) at the homothety points
$\pm\gamma_L(\sigma)$. Then taking into account the explicit form
of the eigenfunctions $\phi_{k=0,1}^L(x)$ and (\ref{chempotent}),
we get that, in contrast to (\ref{loc-dens-cond1}), the
\textit{local condensate} density is
\begin{equation}\label{loc-dens-cond2}
\lim_{L\rightarrow\infty} \sum_{k = 0,1}
\frac{|\phi_k^L(x=\pm\gamma_L(\sigma))|^2}{\e^{\beta(\epsilon_L(k)-
\overline{\mu}_{L}(\beta,\rho))}-1} =  \rho_0(\beta,\rho)
|\sigma|.
\end{equation}

\noindent The same arguments as above show that this condensate
local density varies from \textit{zero} to \textit{infinity}  when
the parameter $a$ in the homothety point positions $\pm\gamma_L(a
\sigma)$ varies in the same interval.
\begin{remark}
These observations can be interpreted  as follows : the
Bose-Einstein condensate for attractive boundary conditions
$\sigma < 0$ is localized in a \textit{logarithmically narrow}
domain in the vicinity of the boundary. In other words this kind of
condensation is a \textit{surface} phenomenon. At the same time
\textit{globally} it is very "visible", since the Kac density
indicates a non-equivalence of ensembles in the presence of the
condensate, see Theorem 3.1.
\end{remark}

\noindent For the generalization to the $\nu$-dimensional case, we
start with the corresponding \textit{local condensate} density:
\begin{equation}\label{loc-dens-cond3}
\rho_{0}(\beta,\rho; \mathbf{x}) : = \lim_{L\rightarrow\infty}
\sum_{\mathbf{k}\in \left\{\mathbb{Z}_{+}^\nu:\, k_{\alpha}= 0,1
\,;\, \alpha = 1,\ldots,\nu \right\}}
\frac{\left|\psi_{\mathbf{k}}^L(\mathbf{x})\right|^2}
{\e^{\beta(E_L(\mathbf{k})-\overline{\mu}_{L}(\beta,\rho))}-1}\,.
\end{equation}
Let $\mathbf{x}$ belong to a bounded domain $D^{\nu}$, containing
the origin of the coordinates $\mathbf{x}=\mathbf{0}$. Then using
the explicit expressions for the eigenfunctions
$\psi_{\mathbf{k}}^L(\mathbf{x})$, see Section 3.1, and by the
same arguments as above for $\nu =1$, we obtain that the limit
(\ref{loc-dens-cond3}) is \textit{zero} for all densities $\rho >
0$.

The \textit{product} structure : $\psi_{\mathbf{k}}^L(\mathbf{x})
= \prod_{i=1}^\nu \phi_{k_i}^L(x_i)$, implies that this
conclusion does not change if we consider instead of
$\mathbf{x}\in D^{\nu}$, the condensate density in the vicinity
of the points corresponding to the shifts where at \textit{least
one} among the $\nu$ arguments remains \textit{unshifted}.\\ On
the other hand, this structure and the asymptotics of $\phi_{k =
0, 1}^L(x)$ for $\left|x\right|\rightarrow \infty$ yields also
that for any $\mathbf{k}\in \left\{\mathbb{Z}_{+}^\nu:
k_{\alpha}= 0,1 \,;\, \alpha = 1,\ldots,\nu \right\}$ one gets
\begin{equation}\label{shift2}
\left(\prod_{\alpha=1}^{\nu}\tau_{\pm\gamma_L(a_{\alpha}\sigma)}
\psi_{\mathbf{k}}^L\right) (\mathbf{x}=0)=
|\sigma|^{\nu/2}L^{-\frac{1}{2}\left(a_{1}^{-1} + a_{2}^{-1}
+\ldots+ a_{\nu}^{-1}\right)} + o
\left(L^{-\frac{1}{2}\left(a_{1}^{-1} + a_{2}^{-1} +\ldots+
a_{\nu}^{-1}\right)}\right)
\end{equation}
as $L\rightarrow\infty$. Then, by virtue of (\ref{chempotent}), the
limit for the local condensate density becomes \textit{non-trivial}:
\begin{equation}\label{loc-dens-cond4}
\lim_{L\rightarrow\infty} \sum_{\mathbf{k}\in
\left\{\mathbb{Z}_{+}^\nu: k_{\alpha}= 0,1 \,;\, \alpha =
1,\ldots,\nu \right\}}
\frac{\left|\left(\prod_{\alpha=1}^{\nu}\tau_{\pm\gamma_L(a_{\alpha}\sigma)}
\psi_{\mathbf{k}}^L\right)
(\mathbf{x}=0)\right|^2}
{\e^{\beta(E_L(\mathbf{k})-\overline{\mu}_{L}(\beta,\rho))}-1} =
\left|\sigma\right|^\nu(\rho - \rho_{c}(\beta)) > 0\, ,
\end{equation} if and only if
\begin{equation}\label{condition1}
a_{1}^{-1} + a_{2}^{-1} +\ldots+ a_{\nu}^{-1} = \nu \,.
\end{equation}
This means that the condensate (up to logarithmic deviations) is
localized essentially in the \textit{corners} of the hypercube
$\Lambda_{L}^{\nu}$, where $L\rightarrow\infty$. We proved
the following statement:
\begin{theorem}\label{condens-loc1}
Let $\mathbf{x}$ be in a bounded domain $D^{\nu}$, containing the
origin of the coordinates $\mathbf{x}=\mathbf{0}$, then the
thermodynamic limit of the local particle density is
\begin{equation}\label{loc-dens-cond5}
\rho(\beta,\rho;\, \mathbf{x}) : = \lim_{L\rightarrow\infty}
\sum_{\mathbf{k}\in \mathbb{Z}_{+}^\nu}
\frac{\left|\psi_{\mathbf{k}}^L(\mathbf{x})\right|^2}
{\e^{\beta(E_L(\mathbf{k})-\overline{\mu}_{L}(\beta,\rho))}-1} =
\frac{1}{\pi^{\nu}}\int_{\mathbb{R}_{+}^\nu}\,d\mathbf{k}
\frac{1}{\e^{\beta (\mathbf{k}^2- \overline{\mu}(\beta,\rho))}-1} ,
\end{equation}
where $\overline{\mu}(\beta,\rho)< - \nu
\sigma^2$ for $\rho < \rho_c(\beta)$ and $\overline{\mu}(\beta,\rho)= -
\nu \sigma^2$ for $\rho \geq \rho_c(\beta)$. Thus  $\rho(\beta,\rho;\,\mathbf{x})= 
\rho_{c}(\beta)$ for $\rho \geq \rho_c(\beta)$, i.e. the local condensate density
$\rho_0(\beta,\rho;\, \mathbf{x})= 0$ for any $\rho > 0$. Whereas
at the homothety points corresponding to the shifts
$\prod_{\alpha=1}^{\nu}\tau_{\pm\gamma_L(a_{\alpha}\sigma)}$ with parameters
satisfying (\ref{condition1}), the local condensate density
(\ref{loc-dens-cond4}) is nontrivial. Moreover, besides being
inhomogeneous it is also anisotropic and essentially localized in the
directions of the corners of the hypercube
$\Lambda_{L\rightarrow\infty}^{\nu}$. Varying the parameters
$\left\{a_{\alpha}\right\}_{\alpha=1}^{\nu}$ in the range
$\left(0, +\infty \right)$ one finds this local condensate
density varying from zero to infinity.
\end{theorem}


\noindent Now we extend Theorem \ref{stell1} on the basis of our
discussion above of the condensate localisation and Theorem
\ref{condens-loc1}. Similar to the one-dimensional case, see
Remark \ref{log-tuning1}, our relevant localized observable in the
$\nu$-dimensional case will be a function
$f\in\mathcal{C}_0^\infty(\mathbb{R}^\nu)$ such that
$supp(f)=(-\delta_1,\delta_1) \times (-\delta_2,\delta_2) \times
\cdots \times (-\delta_\nu , \delta_\nu)\subset \Lambda_L^\nu$ and
with $\delta = \max_{i=1,\ldots,\nu}\delta_i$ such that $\delta <
(\ln L)/2|\sigma|$. Consider in each coordinate the shift over a
distance $\gamma_L(\sigma)=L/2-(2|\sigma|)^{-1}\ln L$:
\begin{equation}
\left( \prod_{\alpha =1}^\nu \tau_{\gamma_L(\sigma)}f
\right)(\mathbf{x}) = f\left(x_1-(L/2-(2|\sigma|)^{-1}\ln
L),\ldots , x_\nu-(L/2-(2|\sigma|)^{-1}\ln L) \right),\nonumber
\end{equation}
then $\prod_{\alpha =1}^\nu \tau_{\gamma_L(\sigma)}f \in
\mathcal{C}_0^\infty(\Lambda_L^\nu)$.

To get the generating functional in the $\nu$-dimensional case,
we compute the limit of the corresponding two-point correlation
function:
\begin{eqnarray}
\lefteqn{\lim_{L\rightarrow\infty}
\omega_{L,\beta,\overline{\mu}_L(\beta,\rho)}^{g.\,c.}
\left(a^\ast\left(\prod_{\alpha=1}^{\nu}\tau_{\gamma_L(\sigma)}f\right)
a\left(\prod_{\alpha=1}^{\nu}\tau_{\gamma_L(\sigma)}f\right)\right)}\nonumber\\
& = & \lim_{L\rightarrow\infty}\sum_{\mathbf{k} \in
\mathbb{Z}_{+}^{\nu}} \left|\prod_{\alpha =1
}^\nu(\hat{f}_{\tau_{\gamma_L(\sigma)}})(\mathbf{k})\right|^2
\frac{1}{\e^{\beta(E_{L}(\mathbf{k})-
\overline{\mu}_L(\beta,\rho))}-1} \nonumber\\ & = &
\lim_{L\rightarrow\infty}\left(\sum_{\mathbf{k}\in
\{\mathbb{Z}_+^\nu : k_\alpha =0,1;\alpha =1,\ldots,\nu\}}
\left|\prod_{\alpha =1
}^\nu(\hat{f}_{\tau_{\gamma_L(\sigma)}})(\mathbf{k})\right|^2
\frac{1}{\e^{\beta(E_{L}(\mathbf{k})-
\overline{\mu}_L(\beta,\rho))}-1}\right.\nonumber\\ & & + \left.
\sum_{n=1}^\infty
\e^{n\beta\overline{\mu}_L(\beta,\rho)}\sum_{\mathbf{k}\in
\mathbb{Z}_+^\nu \setminus \{\mathbb{Z}_+^\nu :\, k_\alpha
=0,1;\,\alpha =1,\, \ldots,\, \nu\}} \e^{-n\beta E_L(\mathbf{k})}
\left|\prod_{\alpha =1
}^\nu(\hat{f}_{\tau_{\gamma_L(\sigma)}})(\mathbf{k})\right|^2\right).
\label{limit-dim3}
\end{eqnarray}
This thermodynamic limit depends on the homothety point
corresponding to the shifts $\prod_{\alpha=1}^{\nu}\tau_{\pm\gamma_L(a_{\alpha}\sigma)}$
with parameters
$a_{\alpha}=1$. Notice that the factor $\left|\prod_{\alpha=1}^\nu
\hat{f}_{\tau_{\gamma_L(\sigma)}}(\mathbf{k}) \right|^2$ is of
the order $O(L^{-\nu})$ for large $L$:
\begin{eqnarray}
\left|\prod_{\alpha=1}^\nu
\hat{f}_{\tau_{\gamma_L(\sigma)}}(\mathbf{k}) \right|^2 & =
& \left|
\int_{\gamma_L(\sigma)-\delta_1}^{\gamma_L(\sigma)+\delta_1}dx_1
\ldots
\int_{\gamma_L(\sigma)-\delta_\nu}^{\gamma_L(\sigma)+\delta_\nu}dx_\nu
\prod_{\alpha=1}^\nu (\tau_{\gamma_L(\sigma)}f)(\mathbf{x})
\overline{\psi_\mathbf{k}^L(\mathbf{x})} |\right|^2 \nonumber \\
 & = & |\sigma|^\nu L^{-\nu}\left|\int_{supp(f)}d\mathbf{x}\,
 f(\mathbf{x}) \prod_{\alpha=1}^\nu \e^{|\sigma|x_\alpha}\right|^2
 + o(L^{-\nu}),
\end{eqnarray}
and for any $\mathbf{k}\in \{\mathbb{Z}_+^\nu: k_\alpha =
0,1;\alpha=1,\ldots ,\nu\}$. Hence, by the same reasoning, which
implies (\ref{loc-dens-cond4}), the first $2^\nu$ terms in
(\ref{limit-dim3}) give:
\begin{eqnarray}
\lefteqn{\lim_{L\rightarrow\infty}\sum_{\mathbf{k}\in
\{\mathbb{Z}_+^\nu : k_\alpha =0,1;\alpha =1,\ldots,\nu\}}
\left|\prod_{\alpha =1
}^\nu(\hat{f}_{\tau_{\gamma_L(\sigma)}})(\mathbf{k})\right|^2
\frac{1}{\e^{\beta(E_{L}(\mathbf{k})-
\overline{\mu}_L(\beta,\rho))}-1}} \nonumber\\ & = &
\rho_0(\beta,\rho) |\sigma|^\nu
\left|\int_{\mathbb{R}^\nu}d\mathbf{x}\,
f(\mathbf{x})\prod_{i=1}^\nu \e^{|\sigma|x_i}\right|^2.
\label{2nuterms}
\end{eqnarray}
For the last term in (\ref{limit-dim3}), we perform the
computations as in Theorem \ref{stell1}, yielding:
\begin{eqnarray}
\lefteqn{\lim_{L\rightarrow\infty}\sum_{\mathbf{k}\in
\mathbb{Z}_+^\nu \setminus \{\mathbb{Z}_+^\nu :\, k_\alpha
=0,1;\,\alpha =1,\, \ldots,\, \nu\}}^\infty \left|\prod_{\alpha=1}^\nu
\hat f_{\tau_{\gamma_L(\sigma)}}(\mathbf{k})\right|^2
\frac{1}{\e^{\beta(E_L(\mathbf{k})-
\overline{\mu}_{L}(\beta,\rho))}-1} } \nonumber \\
& = & \sum_{n=1}^\infty \e^{-n\nu\beta\sigma^2}(4\pi
n\beta)^{-\nu/2}\int_{\mathbb{R}^\nu} d\mathbf{x}
\int_{\mathbb{R}^\nu} d\mathbf{y}
f(\mathbf{x})\overline{f(\mathbf{y})}
\exp\left\{-\frac{\|\mathbf{x}-\mathbf{y}\|^2}{4n\beta}\right\}.\label{term3}
\end{eqnarray}
So, taking the thermodynamic limit $L\rightarrow\infty$ at one of
the homothety points
$\left\{\pm\gamma_L(\sigma)\right\}_{\alpha=1}^\nu$, we get now
the generating functional for $\rho\geq\rho_c(\beta)$:
\begin{eqnarray}
\lefteqn{\lim_{L\rightarrow\infty}\omega_{L,\beta,\overline{\mu}_L(\beta,\rho)}^{g.\,c.}
\left(\prod_{\alpha=1}^\nu W(\tau_{\pm
\gamma(\sigma)}f)\right)}\nonumber\\ & = &
\exp\left(-\frac{1}{4}(f,f)\right)\exp\left(-\frac{1}{2}C^{\pm}
_{\sigma}(\beta,\rho)(f) -\frac{1}{2}(f,g_\sigma (\beta,\rho)
f)\right)\label{genfuncshift}
\end{eqnarray}
with
\begin{eqnarray}
C^{\pm}_{\sigma}(\beta,\rho)(f) & = & \rho_0(\beta,\rho)
|\sigma|^\nu \left|\int_{\mathbb{R}^\nu}d\mathbf{x}\,
f(\mathbf{x})\prod_{\alpha=1}^\nu \e^{\pm
|\sigma|x_\alpha}\right|^2\label{Cnudim}
\end{eqnarray}

\begin{remark}
Again this result is due to a fine (logarithmic) \textit{tuning}
of the position of the homothety points
$\left\{\pm\gamma_L(\sigma)\right\}_{\alpha=1}^\nu$ in the corner
directions of the hypercube $\Lambda_L^\nu$. Indeed, take as in
Theorem \ref{condens-loc1} the shifts $\prod_{\alpha=1}^\nu
\tau_{\pm\gamma_L(a_\alpha \sigma)}$, with $\sum_{\alpha=1}^\nu \left(a_\alpha\right)^{-1} 
> \nu$, i.e. the
homothety points are \textit{more distant} from the corners of
the hypercube. Taking into account the explicit form of the
eigenfunctions for $\mathbf{k} \in \{\mathbb{Z}^\nu_+: k_\alpha
=0,1; \alpha=1,\ldots , \nu\}$, one finds now that
$|\prod_{\alpha=1}^\nu
\hat{f}_{\tau_{\gamma_{L}(a_{\alpha}\sigma)}}(\mathbf{k})|^2$ with $\mathbf{k} \in
\{\mathbb{Z}^\nu_+: k_\alpha =0,1; \alpha=1,\ldots , \nu\}$ goes
like $L^{-(a_1^{-1}+a_2^{-1}+\ldots+a_\nu^{-1})}$ for large $L$.
This implies that the limits of the first $2^\nu$ terms
(\ref{2nuterms}), and hence (\ref{Cnudim}), vanish. So, the
generating functional (\ref{genfuncshift}) has the same form as
for the thermodynamic limit with the homothety point at the
\textit{origin} $\mathbf{x}= \mathbf{0}$. In contrast to that,
the choice $0 < \sum_{\alpha=1}^\nu \left(a_\alpha\right)^{-1} < \nu$
means that the homothety points are too \textit{close} to the
corners of the hypercube $\Lambda_L^\nu$. Then
$|\prod_{\alpha =}^\nu \hat{f}_{\tau_{\pm\gamma_L(a_{\alpha}\sigma)}}(\mathbf{k})|^2$ with $\mathbf{k} \in
\{\mathbb{Z}^\nu_+: k_\alpha =0,1; \alpha=1,\ldots , \nu\}$ goes
to zero slower then $L^{-\nu}$. This implies that the limit
(\ref{Cnudim}) becomes infinite. So, for $\rho \geq
 \rho_{c}(\beta)$ the generating functional (\ref{genfuncshift})
is zero, whereas for $\rho < \rho_{c}(\beta)$ it is nontrivial
with $C^{\pm}_{\sigma}(f) = 0$.
\end{remark}
\noindent Therefore, we proved the following theorem:
\begin{theorem}\label{gen-funct-d3}
The generating functional
$\lim_{L\rightarrow\infty}\omega_{L,\beta,\overline{\mu}_L(\beta,\rho)}^{g.\,c.}
\left(\prod_{\alpha=1}^\nu W(\tau_{\pm\gamma(a_\alpha\sigma)}f)\right)$
on $C_0^{\infty}(\mathbb{R}^{\nu})$ is given by
\begin{eqnarray}
\lefteqn{\lim_{L\rightarrow\infty}\omega_{L,\beta,\overline{\mu}_L(\beta,\rho)}^{g.\,c.}
\left(\prod_{\alpha=1}^\nu
W(\tau_{\pm\gamma(a_\alpha\sigma)}f)\right)}\nonumber\\ & = &
\exp\left(-\frac{1}{4}(f,f)\right)\exp\left(-\frac{1}{2}C^{\pm}
_{\mathbf{a}\sigma}(\beta,\rho)(f) -\frac{1}{2}(f,g_\sigma
(\beta,\rho) f)\right), \label{genfuncshift-1}
\end{eqnarray}
with
\begin{eqnarray}
C^{\pm}_{\mathbf{a}\sigma}(\beta,\rho)(f) & = & \rho_0(\beta,\rho)
|\sigma|^\nu \left|\int_{\mathbb{R}^\nu}d\mathbf{x}\,
f(\mathbf{x})\prod_{i=1}^\nu \e^{\pm
|\sigma|x_i}\right|^2 \chi_{\nu}(\mathbf{a}),\nonumber\\
(g_\sigma(\beta,\rho)
f)(\mathbf{x}) & = & \int_{\mathbb{R}^\nu}d\mathbf{y}\
G_\sigma(\beta,\rho)(\|\mathbf{x}-\mathbf{y}\|)f(\mathbf{y}),\nonumber\\
G_\sigma(\beta,\rho)(r) & = &
(4\pi\beta)^{-\nu/2}\sum_{n=1}^\infty \e^{-r^2/4n\beta}
\frac{\e^{-n\beta\nu\sigma^2}}{n^{\nu/2}}.\nonumber
\end{eqnarray}
Here $\overline{\mu}(\beta,\rho) < -\nu\sigma^2$ for $\rho < \rho_c(\beta)$
and $\overline{\mu}(\beta,\rho) =
-\nu\sigma^2$ for $\rho \geq \rho_c(\beta)$ are the limiting solutions of the 
grand canonical density equation (\ref{g.c.density}), $\rho_0(\beta,\rho)=0$ 
for $\rho < \rho_c(\beta)$ whereas 
$\rho_0(\beta,\rho)= \rho - \rho_c(\beta)$ for $\rho \geq \rho_c(\beta)$
and the function $\chi_{\nu}(\mathbf{a})=
0, 1, +\infty$, respectively for $\nu < \sum_{\alpha=1}^\nu
\left(a_\alpha\right)^{-1}$, $\nu = \sum_{\alpha=1}^\nu
\left(a_\alpha\right)^{-1}$, and $\nu > \sum_{\alpha=1}^\nu
\left(a_\alpha\right)^{-1}$. \hfill $\Box$
\end{theorem}


\section{Concluding remarks}

The main results of our analysis for the free Bose gas with
attractive boundary conditions are contained in the Theorems
\ref{kac}, \ref{stell1} and \ref{gen-funct-d3}.

In Theorem \ref{kac}, we obtain a Kac density function showing
non-equivalence of the canonical and the  
grand canonical ensemble
in the presence of the condensate even if the condensate density is locally zero.

We learn from Theorem \ref{stell1} that the condensation
is not visible in the expectation values of strictly localized
observables because the Bose condensate is situated near the
boundary of an "infinite container". Nevertheless one should observe
the effects of condensation in the equilibrium states, i.e., in
the generating functional.

Theorem \ref{gen-funct-d3} yields the answer. We make precise for
which type of observables the equilibrium states show their
dependence on the condensate. This completes the rigorous analysis
of the problem of (non-)equivalence of ensembles for the free Bose
gas with attractive boundary conditions and the inhomogeneous condensate
localization.

Finally we repeat that we analyzed only the problem taking the
thermodynamic limits in the sense of homothetycally increasing
cubes, with the consequence that the condensate is situated
anisotropically in the direction of the corners of these cubes and is
"localized at infinity". To prove this we tune the homothety point
position at the logarithmic (in the units of the cube size)
distance from the cube boundary. If instead one looks for the
limit of spherical containers, this anisotropy in the positioning
of the condensate should disappear. Does one expect spontaneous
spherical symmetry breaking of the equilibrium states in this
case?
\\
\\
\\
\\
\textbf{Acknowledgements.} The paper was initiated during V.A.Z.'s
visit at the Instituut voor Theoretische Fysica, KU Leuven. He
wishes to thank the Instituut voor Theoretische Fysica for
hospitality.

\end{document}